\documentclass[a4paper,11pt]{article}
\usepackage{jcappub}
\usepackage{graphicx}
\usepackage{subcaption}
\usepackage{dcolumn}
\usepackage{bm}
\usepackage{hyperref} 
\usepackage{orcidlink}
\usepackage[utf8x]{inputenc}

\usepackage{multirow}
\makeatletter
\gdef\@fpheader{Accepted for publication in JCAP}
\makeatother

\newcommand{\gag}{g_{a\gamma}}

\begin{document}

\title{\texorpdfstring{Axion-like Dark Matter Search\\ with Space-based Gravitational Wave Detectors}{Axion-like Dark Matter Search with Space-based Gravitational Wave Detectors}}

\author{Run-Min Yao\orcidlink{0000-0002-7807-6713}}
\emailAdd{yaorunmin@ucas.ac.cn}
\affiliation{School of Fundamental Physics and Mathematical Sciences, Hangzhou Institute for Advanced
Study, UCAS, Hangzhou 310024, China}
\affiliation{University of Chinese Academy of Sciences, Beijing 100049, China}

\author{Xiao-Jun Bi\orcidlink{
0000-0002-5334-9754}}
\emailAdd{bixj@ihep.ac.cn}
\affiliation{University of Chinese Academy of Sciences, Beijing 100049, China}
\affiliation{Key Laboratory of Particle Astrophysics, Institute of High Energy Physics, Chinese Academy of Sciences, Beijing, China}

\author{Peng-Fei Yin\orcidlink{0000-0001-6514-5196}}
\emailAdd{yinpf@ihep.ac.cn}
\affiliation{Key Laboratory of Particle Astrophysics, Institute of High Energy Physics, Chinese Academy of Sciences, Beijing, China}

\author{Qing-Guo Huang\orcidlink{0000-0003-1584-345X}}
\emailAdd{huangqg@itp.ac.cn (corresponding author)}
\affiliation{School of Fundamental Physics and Mathematical Sciences, Hangzhou Institute for Advanced
Study, UCAS, Hangzhou 310024, China}
\affiliation{University of Chinese Academy of Sciences, Beijing 100049, China}
\affiliation{Institute of Theoretical Physics, Chinese Academy
of Sciences}

\date{\today}

\abstract{
We propose a novel modification to the optical benches of space-based gravitational wave detectors (SGWDs) to enable the detection of axion-like dark matter (ALDM)-induced birefringence without altering the polarization of inter-spacecraft laser links. Our design introduces an auxiliary interferometer to convert polarization modulation into measurable phase shifts. Analytical expressions for sensitivity to the ALDM-photon coupling are derived for various time-delay interferometry (TDI) combinations. Projected sensitivity curves demonstrate complementary coverage across the ALDM mass range $10^{-19}\sim10^{-14}\mathrm{eV}$. This approach preserves the original interferometric stability while enabling new physics capabilities for SGWDs.}

\maketitle

\section{\label{sec:introduction}Introduction}
The endeavor to unravel the nature of dark matter, one of the most profound and enduring mysteries in modern physics \cite{Zwicky:1933gu,Bertone:2018krk}, has driven scientists to explore a wide range of theoretical candidates, ranging from ultra-light scalar particles to primordial black holes. Among these possibilities, axion-like dark matter (ALDM) has emerged as a particularly compelling candidate \cite{Preskill:1982cy,Abbott:1982af,Dine:1982ah,Peccei:2006as,Marsh:2015xka,Adams:2022pbo,OHare:2024nmr}, distinguished by a unique set of properties that make it an attractive focus for both theoretical and experimental investigations \cite{Arvanitaki:2009fg,Graham:2015ouw,DiLuzio:2020wdo}. These particles are postulated to possess an extraordinarily low mass, exhibit exceedingly feeble interactions, and exist in large abundance.

One of the most intriguing aspects of ALDM is its interaction with electromagnetic fields, which gives rise to distinctive signatures that are potentially detectable. A key phenomenon associated with ALDM is the birefringence effect \cite{Carroll:1989vb,Carroll:1998zi,Harari:1992ea}, which arises from the coupling between the ALDM field and photons. In the presence of an ALDM field, the phase velocities of left-handed and right-handed circularly polarized light differ while the light propagates through space. This difference causes linearly polarized light to undergo a rotation of its polarization plane as it propagates. This effect provides a distinctive and measurable signature of ALDM. By studying this birefringence effect, researchers can probe the strength of the ALDM-photon coupling, offering a powerful tool to test the existence of these elusive particles and constrain their intrinsic properties. 

While traditional astrophysical observations, such as cosmic microwave background (CMB) polarization studies \cite{Harari:1992ea,Fedderke:2019ajk,Fujita:2020aqt} and celestial polarimetry \cite{diSeregoAlighieri:2010ejm,Fujita:2018zaj,Ivanov:2018byi,Liu:2019brz,Poddar:2020qft,Chen:2021lvo,Yuan:2020xui,Liu:2021zlt,Gan:2023swl}, have been employed to search for birefringence effects induced by ALDM, their sensitivity is ultimately limited by systematic uncertainties and astrophysical foregrounds. To achieve greater precision, new experimental approaches leveraging high-precision interferometry have been proposed. In particular, optical cavities and laser interferometers \cite{DeRocco:2018jwe,Obata:2018vvr,Liu:2018icu,Nagano:2019rbw,Michimura:2019qxr,Heinze:2023nfb}, which can measure tiny phase shifts with extraordinary accuracy, present promising avenues for detecting the subtle imprint of ALDM-induced birefringence. 

Gravitational wave (GW) detectors are well-developed interferometers for probing the phase signature induced by ALDM. The detection of GW has revolutionized astrophysics since the first observation by the LIGO-Virgo collaboration in 2015 \cite{LIGOScientific:2016aoc}, opening a novel observational window into the cosmos. Ground-based interferometers, such as LIGO \cite{LIGOScientific:2014pky}, Virgo \cite{VIRGO:2014yos}, and KAGRA \cite{KAGRA:2018plz}, now routinely detect high-frequency GWs ($10\,\mathrm{Hz}\sim10\,\mathrm{kHz}$) originating from compact binary mergers, including black holes and neutron stars, while pulsar timing arrays \cite{McLaughlin:2013ira,Manchester:2012za,Kramer:2013kea,Manchester:2013ndt} probe the nanohertz regime in search of stochastic signals \cite{NANOGrav:2023gor,Reardon:2023gzh,EPTA:2023fyk,Xu:2023wog} from supermassive black hole binaries. However, the millihertz frequency band—home to mergers of massive black hole binaries, extreme mass-ratio inspirals, and Galactic compact binaries—remains inaccessible to current ground-based or PTA efforts. This gap will be addressed by future space-based detectors, such as LISA \cite{LISA:2024hlh}, Taiji \cite{Luo:2019zal}, TianQin \cite{TianQin:2015yph}, and Big-Bang Observer (BBO) \cite{Crowder:2005nr}, which utilize laser interferometry between spacecraft separated by millions of kilometers to achieve unprecedented phase sensitivity.
These missions employ advanced technologies, including picometer-level metrology, drag-free spacecraft, and time-delay interferometry (TDI) \cite{Tinto:2020fcc}, to suppress noise sources, thereby enabling not only GW astronomy but also synergies with fundamental physics.

Given the exceptional phase sensitivity of SGWDs, typically at the picometer level over million-kilometer baselines, these instruments hold significant potential for detecting the subtle effects induced by the ALDM. Lately, modifications to these SGWDs have been proposed to probe ALDM \cite{Gue:2024txz,Yao:2024hap}. These studies suggest altering the polarization of the laser beam exchanged between the spacecraft (S/C) from linear to circular. With these modifications, the phase difference induced by ALDM would be integrated into the phase measurement of SGWDs. However, such modifications may introduce some issues like the susceptibility to roll rotations of the S/C \cite{muller2017design} and the generation of backreflection \cite{LISA2000ftr}.

In this study, we propose an alternative approach that preserves linear polarization while adding a dedicated auxiliary interferometer on each optical bench (OB). This supplementary interferometer is designed to convert small polarization rotations into a differential phase signal, that can be measured using the established heterodyne detection schemes already implemented for gravitational wave observations. Our approach offers several advantages, including compatibility with current OB designs, the use of established phase measurement techniques with picometer-level sensitivity, and simultaneous operation with gravitational wave detection.

The paper is organized as follows. In Section~\ref{sec:birefringence} we review the ALDM-induced birefringence effect and its impact on the polarization state. Section~\ref{sec:modification} introduces our proposed modification to the optical bench design. Section~\ref{sec:sensitivity} presents a detailed sensitivity analysis of the ALDM-photon coupling. Section~\ref{sec:discussion} discusses practical considerations. Finally, the conclusion is given in Section~\ref{sec:conclusion}. Throughout this paper, all equations are expressed in Lorentz–Heaviside units.

\section{\label{sec:birefringence}ALDM-induced Birefringence}
In this section, we briefly introduce the birefringence effect induced by ALDM and its implications for space-based interferometry. The ALDM-photon interaction is described by the Lagrangian \cite{Carroll:1989vb,Carroll:1991zs,Harari:1992ea}
\begin{equation}
    \mathcal{L}=-\frac{1}{4}F_{\mu\nu}F^{\mu\nu} + \frac{1}{2} \partial^{\mu} a \partial_{\mu} a - \frac{1}{2} m_{a}^2 a^2 -\frac{1}{4} \gag
    aF_{\mu\nu}\tilde{F}^{\mu\nu},
  \label{eq:lagrangian}
\end{equation}
where $a$ is the ALDM field with mass $m_a$, $\gag$ is the ALDM-photon coupling, and $F^{\mu\nu}$ and $\tilde{F}^{\mu\nu}$ are the electromagnetic stress tensor and its dual, respectively.

While photons propagate through the ALDM field, the dispersion relations for the left and the right circular polarization modes attain opposite corrections due to the temporal and spatial variations of the ALDM field, which can be expressed as
\begin{equation}
    \omega_\pm\simeq k\pm\gag\left(\frac{\partial a}{\partial t}+\nabla a\cdot\boldsymbol{\hat{k}}\right)=k\pm\gag n^\mu\nabla_\mu a,
    \label{eq:dispersion}
\end{equation}
where $n^\mu$ denotes the null tangent vector to the photon path.\footnote{The simplified expression found in other works is a special case of Eq. \eqref{eq:dispersion}, valid when the spatial gradient term \(\nabla a \cdot \hat{\mathbf{k}}\) is negligible. This assumption is valid since we are considering the coherence length of the ALDM field significantly exceeds the distance between S/C.} This effect is known as ALDM-induced birefringence.

The differential dispersion relation implies that left and right circularly polarized waves propagate with slightly different phase velocities, leading to a cumulative phase difference between them. For linearly polarized light, which can be decomposed into equal amplitudes of left and right circular polarizations, this phase difference manifests as a rotation of the polarization plane.

The phase difference between the left and right circular polarization can be probed by SGWDs, constituting the central detection principle proposed in \cite{Gue:2024txz,Yao:2024hap}. In their proposals, the linearly polarized light transmitted between S/C is substituted with circularly polarized light, and the phase shift induced by the birefringence effect is integrated into the interferometer signal. However, although the current design of SGWDs utilizes linearly polarized laser links, the past designs of SGWDs favored circularly polarized light for the inter-satellite path \cite{muller2017design}, and both options were under consideration \cite{ESA2011LISAYellowBook}. It seems that the advantages of linear polarization were deemed more favorable after detailed trade-off studies. Using circular polarization requires converting from linear to circular via quarter-wave plates and then back to linear at the receiving end. Extra conversion steps are sensitive to relative rotations, coating phase shifts, and alignment errors between spacecraft. Circular polarization in phase sensitive ranging yields a susceptibility to roll rotations of the S/C \cite{muller2017design}. Besides, optical components, such as quarter-wave plates, are main sources of backreflection, and it is easier to design and to predict the behaviour of the antireflection coatings of the optics working only with linear polarization \cite{LISA2000ftr}. 
This motivates the exploration of alternative approaches that preserve the primary light path of SGWDs.

For linearly polarized light, the differential shift leads to a rotation of the electric vector position angle, expressed as:
\begin{equation}
\begin{aligned}
    \Delta \phi_a = &\frac{1}{2}\int_{t_\mathrm{emit}}^{t_\mathrm{obs}}(\omega_+-\omega_-)dt\\
    = & \frac{1}{2}\gag\left[a(x_\mathrm{obs}^\mu)-a(x_\mathrm{emit}^\mu)\right].
\end{aligned}
\end{equation}
This rotation only depends on the difference in the ALDM field values at emission and observation points, denoted by $x_\mathrm{emit}^\mu$ and $x_\mathrm{obs}^\mu$. 

Neglecting the back-action term of light on the ALDM, the wave function of the ALDM field, derived from the equations of motion arising from Eq. \eqref{eq:lagrangian}, can be written as
\begin{equation}
    a(t,\boldsymbol{x})=a_0 \cos(\omega_at-\boldsymbol{k}_a\cdot\boldsymbol{x}+\Phi),
    \label{eq:ALDM_field}
\end{equation}
where $a_0$ is the field amplitude, related to the dark matter density $\rho_\mathrm{DM}=m_a^2a_0^2/2\approx0.4\:\mathrm{GeV\cdot cm^{-3}}$ \cite{McMillan:2011wd}, and $\Phi$ denotes an unknown phase. 
In the non-relativistic limit, the frequency $\omega_a$ and wave vector $\boldsymbol{k}_a$ are given by $\omega_a\simeq m_a(1+\boldsymbol{v}^2/2)$ and $\boldsymbol{k}_a\simeq m_a\boldsymbol{v}$, respectively, where $\boldsymbol{v}$ is the collective velocity of the ALDM with $v\sim10^{-3}$. When the coherence length of the ALDM field, $\lambda_a=2\pi/k_a$, significantly exceeds the distance between S/C, denoted as $L$, Eq. \eqref{eq:ALDM_field} can be approximated as
\begin{equation}
    a(t)\approx\frac{\sqrt{2\rho_\mathrm{DM}}}{m_a}\cos(m_at+\Phi).
\end{equation}
This approximation is valid for the majority of the ALDM mass range of interest ($10^{-19}$ to $10^{-14}$ eV), as the corresponding coherence length exceeds the typical arm length of SGWDs, which is on the order of $\sim10^9\:\mathrm{m}$. Under these conditions, the ALDM field can be treated as temporally oscillating but spatially uniform across the detector network.

The expected change in the polarization angle induced by the birefringence effect for a single-link light reads
\begin{equation}
\begin{aligned}
    \Delta\phi_a=&\frac{1}{2}\gag\frac{\sqrt{2\rho_\mathrm{DM}}}{m_a}\left\{\cos(m_at+\Phi)-\cos\left[m_a(t-L)+\Phi\right]\right\}\\
    =&-\gag\frac{\sqrt{2\rho_\mathrm{DM}}}{m_a}\sin\frac{m_aL}{2}\sin\left(m_at+\Phi-\frac{m_aL}{2}\right)\\
    \sim&-10^{-7}\left(\frac{\gag}{10^{-11}\,\mathrm{GeV}^{-1}}\right)\left(\frac{\rho_\mathrm{DM}}{0.4\,\mathrm{GeV}\cdot \mathrm{cm}^{-3}}\right)^{1/2}\left(\frac{10^{-16}\,\mathrm{eV}}{m_a}\right)\\
    &\times\sin\left(\frac{m_aL}{2}\right)\sin\left(m_at+\Phi-\frac{m_aL}{2}\right).
\end{aligned}
\end{equation}    
The numerical estimate in the above equation reveals that the birefringence-induced polarization rotation is quite small, approximately $10^{-7}$ radians for typical parameter values. This underscores the challenge of detecting such a signal and motivates the need for highly sensitive interferometric techniques.

Without loss of generality, we suppose that the polarization angle between the transmitting laser light from remote S/C and the optical bench reference basis is $\theta=\theta_0+\Delta\phi_a$, where $\theta_0$ is the misalignment polarization angle without ALDM.
The received light affected by the ALDM oscillation can be written as
\begin{equation}
    \boldsymbol{E}_\mathrm{RX}\propto\cos\theta \hat{\boldsymbol{s}}+\sin\theta\hat{\boldsymbol{p}}.
    \label{eq:rx1}
\end{equation}
Here $\hat{\boldsymbol{s}}$ and $\hat{\boldsymbol{p}}$ represent the unit vectors in the s-polarization and p-polarization directions, respectively.

In general setup of SGWDs, the angle $\theta_0$ is calibrated to be 0, then
\begin{equation}
    \boldsymbol{E}_\mathrm{RX}\propto\cos\Delta\phi_a \hat{\boldsymbol{s}}+\sin\Delta\phi_a\hat{\boldsymbol{p}}.
    \label{eq:rx2}
\end{equation}
The presence of the ALDM field induces a small p-polarization component proportional to $\sin\Delta\phi_a$, while slightly reducing the s-polarization component proportional to $\cos\Delta\phi_a$. From Eq. \eqref{eq:rx2}, the s-polarized component of the received light undergoes periodic modulation at $\mathcal{O}(\Delta\phi_a^2)$ due to the ALDM field. However, this modulation lies below the sensitivity threshold of the current design of SGWDs. A possible approach to make the interferometer sensitive to the birefringence effect is to measure the amplitude of the p-polarized component of the received light, as proposed in studies such as \cite{Nagano:2019rbw}. Nevertheless, direct detection of the p-polarized signal poses significant challenges in SGWDs, due to the extremely low power of the received light, which is on the order of $\sim700\,\mathrm{pW}$. Our proposal circumvents this limitation by converting the polarization rotation into a differential phase measurement through the implementation of an auxiliary interferometer.

\section{\label{sec:modification}Proposed Modification}
\begin{figure}[htpb]
\includegraphics[width=1.0\columnwidth]{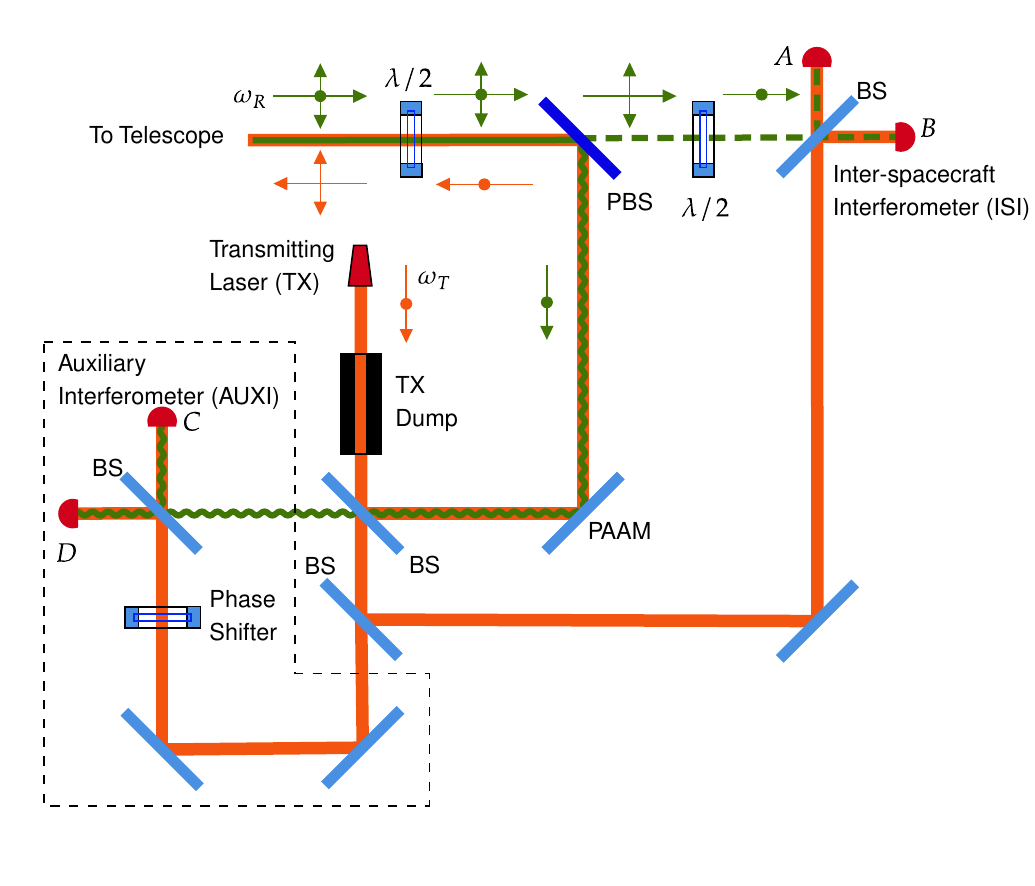}
  \caption{\label{fig:modified_OB} Schematic of the modified optical bench design. The dotted box highlights the additional auxiliary interferometer, phase shifter, and associated optics added to the current OB configuration. The orange beam represents the transmitted light, the green dotted path indicates the received s-polarized light, and the green waving path shows the received p-polarized light. The arrows show the beam direction of propagation, whereas the double-tipped arrows and dots represent the in-plane and out-of-plane polarizations. PBS: polarizing beam splitter; BS: beam splitter; $\lambda/2$: half-wave plate; PAAM: point-ahead angle mechanism; A, B, C, D: photodetector.}.
\end{figure}
In order to measure the ALDM-induced effect without altering the linearly polarized light, a viable solution is to convert power measurement to phase measurement. This can be achieved by introducing an auxiliary interferometer for the p-polarized component of the received light on the OB. In this work, we propose a modification of the OB design to detect the birefringence effect signals.

The main innovation of our approach lies in maintaining the established linear polarization scheme while adding a dedicated detection path for the p-polarization component. This strategy preserves the advantages of the current SGWD design while enhancing sensitivity to new physics.

The proposed modification to the OB is illustrated in Fig. \ref{fig:modified_OB}. The optical elements enclosed within the dotted line, including the auxiliary interferometer, a phase shifter, and associated mirrors, are added to the OB, while the remaining components represent a simplified version of the current OB design for SGWDs. Polarizing beam splitters (PBS) are employed to separate the transmitted and received light. In the schematic, the transmitted light is depicted in orange, the received s-polarized light is represented by green dotted lines, and the received p-polarized light is illustrated with green wavy lines. The proposed optical design is primarily illustrative, and a full engineering design of the optical bench is beyond the scope of this study.

In our proposed design, the received p-polarized component is converted to s-polarized and then directed by the PBS into the auxiliary interferometer path, while the s-polarized component propagates along the standard interferometer path. The auxiliary path includes a phase shifter that enables precise control of the relative phase between the two interferometer outputs, which is crucial for optimizing the detection sensitivity. The TX dump suppresses the back-scattered light from the received p-polarized light.

For SGWDs, the interference between a local transmitted beam and the received beam results in heterodyne signals at the photodiodes. In the presence of the ALDM field, the polarization direction of the received light is $\theta=\theta_0+\Delta\phi_a$. In the inter-spacecraft interferometer (ISI), the received amplitude is modulated by a factor of $\cos\theta$, while in the auxiliary interferometer (AUXI), the modulation appears as $\sin\theta$. With the auxiliary channel, the amplitude modulation due to the polarization angle can be converted into phase signals $\phi_\mathrm{AUXI}$ and $\phi_\mathrm{ISI}$. The phase $\phi_\mathrm{AUXI}$ is read out by combining the heterodyne outputs of the ISI and the auxiliary interferometer appropriately while $\phi_\mathrm{ISI}$ is the phase signal from the ISI. Comparing these two signals, one can recover the polarization angle
\begin{equation}
    \theta\simeq\phi_\mathrm{AUXI} - \phi_\mathrm{ISI},
    \label{eq:phase_difference}
\end{equation}
thereby isolating the ALDM contribution. This differential measurement approach offers a significant advantage: common-mode noises, such as laser frequency fluctuations and S/C motion, are largely canceled out, leaving primarily the ALDM-induced signal.

In the ISI, a fraction of the local transmitted beam $\sqrt{2I_0}e^{i\omega_\mathrm{T} t}$ interferes with the light received from the remote S/C $p_\mathrm{ISI}\sqrt{2I_0}e^{i(\omega_R t+\phi)}$, where $I_{ref}=2I_0$ is the power of the local reference, $p_\mathrm{ISI}^2\sim10^{-7}$ is defined as the ratio of the received optical power to the local reference beam when the incoming light is in a pure s-polarized state, $\omega_\mathrm{T}$ is the angular frequency of the transmitted beam, $\omega_R$ is the angular frequency of the received beam, and $\phi$ is the phase of the received wavefront. In the presence of the ALDM field, the power of the received light in the ISI is modulated by the birefringence effect, expressed as
\begin{equation}
    p_\mathrm{ISI}^\prime = \cos\theta p_\mathrm{ISI}.
\end{equation}
The total field at the photodiodes include both the primary signal and potential noise sources, such as backscattered light, which can introduce spurious signals if not properly mitigated. Its expression can be given by
\begin{gather}
    E_A=\sqrt{2I_0}e^{i\omega_\mathrm{T} t}\left[1+p_A^\prime e^{i(\Omega t+\phi+\pi/2)}+u_Ae^{i(\psi+\pi/2)}\right],\\
    E_B=\sqrt{2I_0}e^{i\omega_\mathrm{T} t}\left[e^{i\pi/2}+p_B^\prime e^{i(\Omega t+\phi)}+u_Be^{i\psi}\right],
\end{gather}
where the subscripts A and B label the diode ports, $\Omega=\omega_R-\omega_\mathrm{T}$ is the heterodyne frequency, $p_\mathrm{ISI}^{\prime2}=p_A^{\prime2}+p_B^{\prime2}\ll1$, the fraction $u_\mathrm{ISI}^2=u_A^2+u_B^2\ll1$ and phase $\psi$ are the fractional power and phase of the backscattered light field.
The heterodyne signals at the photodiodes A and B are
\begin{gather}
    S_A(t)=2I_0\left[1+p_A^{\prime2}+u_A^2-2u_A\sin\psi-2p_A^\prime\sin(\Omega t+\phi)+2p_A^\prime u_A\cos(\Omega t+\phi-\psi)\right],\\
    S_B(t)=2I_0\left[1+p_B^{\prime2}+u_B^2+2u_B\sin\psi+2p_B^\prime\sin(\Omega t+\phi)+2p_B^\prime u_B\cos(\Omega t+\phi-\psi)\right].
\end{gather}
The photodiode outputs contain the desired heterodyne signal at frequency $\Omega$, along with direct current (DC) terms and beat signals between the received and backscattered light.

The digitized signal difference is
\begin{equation}
\begin{aligned} 
S_B(t)-S_A(t)&=2I_0\left[p_B^{\prime2}-p_A^{\prime2}+u_B^2-u_A^2+2(u_B+u_A)\sin\psi\right.\\
&\left.+2(p_B^{\prime}+p_A^{\prime})\sin(\Omega t+\phi)+2(p_B^{\prime}u_B-p_A^{\prime}u_A)\cos(\Omega t+\phi-\psi)\right].
\end{aligned}
\end{equation}
This signal difference is multiplied by a local oscillator with frequency $\Omega$, 
\begin{equation}
\phi_\mathrm{ISI}=\arg\left[\int_0^{2n\pi/\Omega}\left[S_B(t)-S_A(t)\right]e^{-i\Omega t}dt\right],
\end{equation}
where the integration extends over $n$ oscillator cycles. The integrated output is proportional to the phase difference between the signal and oscillator, where only alternating current (AC) terms contribute. This demodulation process effectively extracts the phase information from the heterodyne beat signal, which constitutes the primary quantity measured in the interferometer.

From the integrated output, the phase of the heterodyne-signal can be recovered as \cite{Sasso:2018bau}
\begin{equation}
    \phi_\mathrm{ISI} \approx \phi+\pi/2+\tilde{u}\cos\psi-\tilde{u}^2\sin\psi\cos\psi,
\end{equation}
with 
\begin{equation}
   \tilde{u}_\mathrm{ISI}=\frac{p_A^\prime u_A-p_B^\prime u_B}{p_A^\prime+p_B^\prime}. 
\end{equation}
The terms involving $\tilde{u}_\mathrm{ISI}$ represent corrections due to backscattering effects. Under normal operating conditions, these terms are small and can be calibrated out, but they are included here for completeness.

Similarly, for the AUXI, the signal at the photodiode C and D is given by
\begin{equation}
\begin{aligned} 
S_D-S_C&=2I_1\left[p_D^{\prime2}-p_C^{\prime2}+u_D^2-u_C^2+2(u_D+u_C)\sin\tilde{\psi}\right.\\
&\left.+2(p_D^{\prime}+p_C^{\prime})\sin(\Omega t+\phi+\Delta\phi)+2(p_D^{\prime}u_D-p_C^{\prime}u_C)\cos(\Omega t+\phi+\Delta\phi-\tilde{\psi})\right]
\end{aligned}
\end{equation}
where
\begin{equation}
\Delta\phi=\Delta\phi_\mathrm{Tx}+\Delta\phi_\mathrm{Rx}+\Delta\phi_\mathrm{PS}
\end{equation}
represents the phase shift relative to the ISI. Here, $\Delta\phi_\mathrm{Tx}$ and $\Delta\phi_\mathrm{Rx}$ are the phase differences arising from the path length differences of the transmitted light and received light, respectively. $\Delta\phi_\mathrm{PS}$ is the phase shift introduced by the phase shifter, which can be preset on Earth. The term $I_1$ is the power of the reference beam in the AUXI, and $p_\mathrm{AUXI}^2=p_C^2+p_D^2$ denotes the ratio of the received optical power to the reference beam when the incoming light is in a pure p-polarized state. The parameter $u_\mathrm{AUXI}^2=u_C^2+u_D^2$ denotes the ratio of the power of the backscattered light to the reference beam. Likewise, the power of the received light in the AUXI is modulated by the birefringence effect
\begin{equation}
    p_\mathrm{AUXI}^\prime=\sin\theta p_\mathrm{AUXI}.
\end{equation}

For simplicity, we assume $I_1=I_0$ and $p_\mathrm{AUXI}=p_\mathrm{ISI}$. In addition, the heterodyne signals are considered to be balanced \cite{Carleton:68,Fleddermann:2017zna} with $p_k=p/\sqrt{2},\,k\in\{A,B,C,D\}$. Under these conditions, the combined AC components of the signal differences from both interferometers can be expressed as
\begin{equation}
\begin{aligned}
    &AC\{(S_D-S_C)+(S_B-S_A)\}\\
    =&\sqrt{2}pI_0\Big[\cos\theta\sin(\Omega t+\phi)+\sin\theta\sin(\Omega t+\phi+\Delta\phi)\\
    +&(u_B-u_A)\cos(\Omega t+\phi-\psi)+(u_D-u_C)\cos(\Omega t+\phi+\Delta\phi-\tilde{\psi})\Big]\\
    =&\sqrt{2}pI_0\Big[\sin(\Omega t+\phi+\theta)+\sin\theta[\sin(\Omega t+\phi+\Delta\phi)-\cos(\Omega t+\phi)]\\
    +&(u_B-u_A)\cos(\Omega t+\phi-\psi)+(u_D-u_C)\cos(\Omega t+\phi+\Delta\phi-\tilde{\psi})\Big].
\end{aligned}
\end{equation}
By combining the signals from both interferometers, we create a measurement that is directly sensitive to the ALDM-induced phase shift $\Delta\phi_a$. The algebraic manipulation above demonstrates how the combined signal contains information about both the standard interferometric phase and the ALDM-induced birefringence effect.
Regulating $\Delta\phi=\pi/2$, we obtain
\begin{equation}
\begin{aligned}
    &AC\{(S_D-S_C)+(S_B-S_A)\}\\
    =&\sqrt{2}pI_0\Big[\sin(\Omega t+\phi+\theta)\\
    +&(u_B-u_A)\cos(\Omega t+\phi-\psi)-(u_D-u_C)\sin(\Omega t+\phi-\tilde{\psi})\Big].
\end{aligned}
\end{equation}
The choice of $\Delta\phi=\pi/2$ is a crucial,  as it maximizes the sensitivity to the ALDM signal by ensuring that the auxiliary interferometer operates at the optimal phase difference relative to the main interferometer. This configuration converts the polarization angle into a phase shift relative to $\phi_\mathrm{ISI}$, which can be directly measured by standard phase detection techniques.

The phase of the heterodyne-signal can then be recovered as
\begin{equation}
    \phi_\mathrm{AUXI}\approx \phi+\theta+\pi/2+\phi_b,
\end{equation}
where
\begin{equation}
\phi_b\approx\left[(u_A-u_B)\cos\psi-(u_C-u_D)\sin\tilde{\psi}\right]\left[1-(u_A-u_B)\sin\psi-(u_C-u_D)\cos\tilde{\psi}\right].
\end{equation}
The phase $\phi_\mathrm{AUXI}$ is the key measurand of the AUXI. By subtracting $\phi_\mathrm{ISI}$ from $\phi_\mathrm{AUXI}$, we can recover the polarization angle as expressed in Eq. \eqref{eq:phase_difference}.\footnote{In a more realistic case, we can consider $I_1=f_I^2 I_0$ and $p_\mathrm{AUXI}=f_p p_\mathrm{ISI}=f_p\frac{p}{\sqrt{2}}$, where $f_I$ and $f_p$ are determined by the reflectance and transmittance coefficients of the beam splitters. Then the phase difference would be $\phi_\mathrm{AUXI}-\phi_\mathrm{ISI}\simeq\arctan(\frac{1}{f_If_p}\tan\theta)$.} Then the change in the polarization angle induced by the ALDM for a single-link light can be obtained naturally.

In the current design of SGWDs, the misalignment polarization angle $\theta_0$ is calibrated to be 0. For the modified OB we proposed above, it would be extremely difficult to detect the p-polarized component, given the weak received laser power and the further suppression. To circumvent this issue, an intentional initial misalignment angle needs to be included from the outset, ensuring that both s- and p-components are simultaneously present and measurable. The observable quantity is the modulation of this preexisting polarization angle due to the ALDM-induced birefringence, rather than the appearance of an additional polarization component.

The total measurement noise in the SGWDs mainly comes from three sources: photon (shot) noise, residual laser frequency noise and spacecraft jitter noise. Photon noise dominates at higher frequencies ($>1\,\mathrm{mHz}$); at lower frequencies ($\approx0.1\,\mathrm{mHz}$), other noise sources dominate. The received optical power affects only the photon-noise-limited part of the spectrum. The noise spectral density due to photon noise is $S_\mathrm{shot}\propto \lambda/\sqrt{P_\mathrm{rec}}$, where $\lambda$ denotes the wavelength of the laser and $P_\mathrm{rec}$ denotes the received optical power. Within a 10–50\% reduction of received optical power, SGWDs' science performance would not be seriously affected. The high-frequency photon-noise-limited sensitivity would degrade slightly, but overall mission goals would remain achievable. Therefore, the introduction of the misalignment angle is acceptable and meaningful.

In the subsequent analysis, we assume that the contribution of backscattered light is negligible. This assumption is reasonable, as modern space-based interferometer designs incorporate advanced stray light suppression techniques. Moreover, the influence of backscattered light can be further mitigated through meticulous optical design and sophisticated signal processing methods.

\section{\label{sec:sensitivity}Sensitivity to the ALDM-photon coupling}

In this section, we derive the sensitivity of the modified SGWDs to the ALDM-photon coupling $\gag$. We begin by analyzing the single-arm signal and subsequently extend the analysis to various TDI combinations.

\subsection{\label{subsec:single_arm}Single-arm signal}
By differentiating the phase difference induced by the ALDM-induced birefringence, we define the single-arm response as
\footnote{The single-arm signal seems a little different from the equation in \cite{Gue:2024txz}, this originates from the typo of the definition of the reduced Planck energy in their paper, which is supposed to be $E_P=\sqrt{\hbar c/\kappa}\simeq2.435\times10^{18}\,\mathrm{GeV}$. With the correct definition, the factor $\sqrt{8\pi G}$ would be canceled out.}
\begin{equation}
\begin{aligned}
\eta_{rs} &\equiv \frac{1}{\omega_r}\frac{d\left(\phi_\mathrm{AUXI}-\phi_\mathrm{ISI}\right)}{dt}
\approx \frac{1}{\omega_r}\frac{d\Delta\phi_a}{dt} \\
& = -\frac{\gag}{\omega_r}\sqrt{2\rho_\mathrm{DM}}\sin\frac{m_a L}{2}\cos\Big(m_a t+\Phi-\frac{m_aL}{2}\Big),
\end{aligned}
\end{equation}
and similarly
\begin{equation}
\bar{\eta}_{rs} \equiv \frac{1}{\omega_r}\frac{d\left(\phi_\mathrm{ISI}-\phi_\mathrm{AUXI}\right)}{dt}
\approx -\frac{1}{\omega_r}\frac{d\Delta\phi_a}{dt},
\end{equation}
where $\omega_r$ denotes the laser frequency, and the subscripts $r$ and $s$ represent the indices of the S/C that received and sent the light beam, respectively. The single-arm responses $\eta_{rs}$ and $\bar{\eta}_{rs}$ provide a direct measure of the time derivative of the ALDM-induced phase shift. In practice, SGWDs employ TDI to suppress noises, such as laser frequency noise and OB motion noise, by combining the signals from multiple interferometric arms with appropriate time delays \cite{pub.1058615630,Tinto:1999yr,Dhurandhar:2001tct,Tinto:2020fcc}. This technique enhances sensitivity to gravitational wave signals. From the definition of the single-arm response to the ALDM effect, the subtraction between the two channels cancels out these common-mode noises.

Working with the time derivative of the phase difference, rather than the phase difference itself, offers significant advantages for signal processing. It facilitates the implementation of TDI techniques and enables the separation of the ALDM signal from slowly varying systematic effects. Additionally, the frequency dependence of the signal provides a distinctive signature,  aiding in the discrimination against potential background sources.

\subsection{\label{subsec:TDI_signal}Signals for different TDI combinations}

TDI is implemented in SGWDs to synthesize virtual interferometric channels immune to dominant noise sources while preserving the physical signals of interest. This technique is essential for achieving the sensitivity required for gravitational wave detection, and we can leverage the same approach for ALDM searches without compromising the primary science objectives.

For each TDI combination, the ALDM-induced signal is weighted by a factor that depends on the arm length $L$ and the ALDM mass $m_a$. For example, the Sagnac variable $\alpha$ is given by \cite{pub.1058615630}
\begin{equation}
\begin{aligned}
\alpha(t) &= \Big[\bar{\eta}_{12}+D_{12}\bar{\eta}_{23}+D_{12}D_{23}\bar{\eta}_{31}\Big] - \Big[\eta_{13}+D_{13}\eta_{32}+D_{13}D_{32}\eta_{21}\Big] \\
& = -2\Big(\eta_{13}+D_{13}\eta_{32}+D_{13}D_{32}\eta_{21}\Big)\\
& = 2\,\frac{\gag}{\omega_r}\sqrt{2\rho_\mathrm{DM}}\sin\frac{3m_a L}{2}\cos\Big(m_a t+\Phi-\frac{3m_aL}{2}\Big),
\end{aligned}
\end{equation}
where $D_{ij}$ represents the delay operator corresponding to the light travel time between S/C $i$ and $j$.
Extending this to the second-generation Sagnac variable \cite{Vallisneri:2005ji}, we obtain:
\begin{equation}
\begin{aligned}
\alpha_2(t) &= \Big(1-D_{12}D_{23}D_{31}\Big)\alpha \\
& = 4\,\frac{\gag}{\omega_r}\sqrt{2\rho_\mathrm{DM}}\sin^2\frac{3m_a L}{2}\sin\Big(m_a t+\Phi-\frac{3m_aL}{2}\Big).
\end{aligned}
\end{equation}

While the modifications in \cite{Gue:2024txz, Yao:2024hap} can be applied to various TDI combinations (see, e.g., the Appendix of \cite{Yao:2024hap} and Ref. \cite{Liu:2025hwn}\footnote{A related construction in the circular-polarization setup is also discussed in \cite{Liu:2025hwn}, where the ALDM response is formed from linear combinations rather than from two helicity-defined virtual paths.}), the resulting sensitivity is maximized for Sagnac-type configurations due to the helicity–direction binding. In contrast, our proposed approach provides a distinct framework that avoids modifying the laser polarization state. This allows for optimized sensitivity performance across a broader range of TDI channels, maintaining robustness in sensitivity even in non-Sagnac configurations. For example, the X variable reads
\begin{equation}
\begin{aligned}
X(t) &= \Big(1-D_{13}D_{31}\Big)\Big(\eta_{12}+D_{12}\eta_{21}\Big)
- \Big(1-D_{12}D_{21}\Big)\Big(\bar{\eta}_{13}+D_{13}\bar{\eta}_{31}\Big)\\
& = 4\,\frac{\gag}{\omega_r}\sqrt{2\rho_\mathrm{DM}}\sin^2(m_a L)\sin\Big(m_a t+\Phi-2m_aL\Big),
\end{aligned}
\end{equation}
with its second-generation combination
\begin{equation}
\begin{aligned}
X_2(t) &= \Big(1-D_{31}^{2}D_{12}^{2}\Big)X \\
& = 8\,\frac{\gag}{\omega_r}\sqrt{2\rho_\mathrm{DM}}\sin^2(m_a L)\sin(2m_aL)\cos\Big(m_a t+\Phi-4m_aL\Big).
\end{aligned}
\end{equation}

The fully symmetric $\zeta$ variable is expressed as \cite{Muratore:2021uqj}
\begin{equation}
\begin{aligned}
\zeta(t) &= D_{12}\Big(\bar{\eta}_{31}-\eta_{32}\Big) + D_{23}\Big(\bar{\eta}_{12}-\eta_{13}\Big) + D_{31}\Big(\bar{\eta}_{23}-\eta_{21}\Big)\\
& = 6\,\frac{\gag}{\omega_r}\sqrt{2\rho_\mathrm{DM}}\sin\frac{m_a L}{2}\cos\Big(m_a t+\Phi-\frac{3m_aL}{2}\Big),
\end{aligned}
\end{equation}
with a related second-generation variable \cite{Muratore:2020mdf}
\begin{equation}
\begin{aligned}
\zeta_{2}(t) &= \Big(D_{31}-D_{12}D_{23}\Big)\zeta\\
& = -12\,\frac{\gag}{\omega_r}\sqrt{2\rho_\mathrm{DM}}\sin^2\frac{m_a L}{2}\cos\Big(m_a t+\Phi-3m_aL\Big).
\end{aligned}
\end{equation}

For a signal $a(t)=A\sin(m_at+\phi)$ or $a(t)=A\cos(m_at+\phi)$, its one-sided power spectral density (PSD) is given by 
\begin{equation}
    S(f)=\frac{A^2}{2\pi^2T_{obs}}\frac{\sin^2[\pi(f-f_a)T_{obs}]}{(f-f_a)^2},
\end{equation}
where $f_a=m_a/2\pi$. For each TDI variable $i\in\{\alpha,\alpha_2,X,X_2,\zeta,\zeta_2\}$, the one-sided PSD of the signal is 
\begin{equation}
    S_i(f)=\frac{4C_i^2(f)\gag^2\rho_\mathrm{DM}}{\pi^2\omega_r^2T_\mathrm{obs}}\frac{\sin^2[\pi(f-f_a)T_\mathrm{obs}]}{(f-f_a)^2},
    \label{eq:signal_psd}
\end{equation}
where $f_a=m_a/2\pi$ is the ALDM Compton frequency, and $T_{\mathrm{obs}}$ is the observation time. The factor $C_i(f)$ is a sine term that encapsulates the response of the specific TDI combination
\begin{gather}
    C_X(f)=2\sin^2(2\pi fL),\quad C_{X_2}(f)=4\sin^2(2\pi fL)\sin(4\pi fL),\\
    C_\alpha(f)=\sin(3\pi fL),\quad C_{\alpha_2}(f)=2\sin^2(3\pi fL),\\
    C_\zeta(f)=3\sin(\pi fL),\quad C_{\zeta_2}(f)=6\sin^2(\pi fL).
\end{gather}
In the limit of $f\to f_a$, Eq. \eqref{eq:signal_psd} simplifies to
\begin{equation}
    S_i(f)=4C_i^2(f)\gag^2\rho_\mathrm{DM}T_\mathrm{obs}/\omega_r^2.
\end{equation}

\subsection{\label{subsec:expected_sensitivity}Expected sensitivity}

The two primary noise sources that limit the sensitivity of SGWDs are optical metrology system (OMS) noise and test mass acceleration noise. The OMS noise arises from imperfections in the laser interferometry system used to measure the distance between test masses in different spacecraft, such as residual frequency noise, shot noise, and pathlength noise. The acceleration noise originates from non-gravitational forces acting on the test masses. Both are broadband and stochastic in nature.

Scattered-light noise, originating from parasitic reflections and microscopic surface scattering, is largely incoherent over time: while it interferes coherently within a single optical path, its phase rapidly decorrelates on time scales shorter than a few seconds due to thermal and mechanical perturbations. As a result, its contribution to the long-term phase noise power spectral density remains far below that of the primary noise sources. Provided that the optical layout is not modified in a way that enhances coherent low-frequency scattering paths, such scattered-light noise can be safely neglected in sensitivity estimates dominated by acceleration and readout noise.

In contrast, the ALDM-induced birefringence signal is a temporally coherent, narrowband effect, determined by the difference of the axion field amplitude between the emission and reception points of the laser beam. For the typical SGWDs arm length, the induced polarization rotation is of order $10^{-7}\,\mathrm{rad}$. Although the instantaneous amplitude of this signal is below the phase noise level, its long coherence time enables coherent integration that enhances the signal-to-noise ratio as $T^{1/2}$. Under these conditions, broadband random noise—including residual scattered-light fluctuations—averages down efficiently, while only narrowband or persistent scattering features near the axion Compton frequency could produce correlated artifacts. 

In our analysis, we assume that the proposed modifications to the OB do not significantly alter the noise PSDs. For completeness, we note that the noise PSDs for the $X$, $\alpha$, and $\zeta$ channels, as well as their second-generation counterparts, are given by expressions of the form \cite{Hartwig:2023pft}\footnote{Here we assume that the OMS and acceleration contributions for the Doppler variables $\eta$ and $\bar{\eta}$ are uncorrelated.}
\begin{align}
N_{X}(f)&=16\sin^2(2\pi fL)\{[3+\cos(4\pi fL)]S_\mathrm{acc}(f)+S_\mathrm{oms}(f)\},\\
N_{X_2}(f)&=4\sin^2(4\pi fL)\,N_{X}(f),\\
N_{\alpha}(f)&=8\Big[2\sin^2(\pi fL)+\sin^2(3\pi fL)\Big]S_\mathrm{acc}(f)+6S_\mathrm{oms}(f),\\
N_{\alpha_2}(f)&=4\sin^2(3\pi fL)\,N_{\alpha}(f),\\
N_{\zeta}(f)&=6\Big[4\sin^2(\pi fL)S_\mathrm{acc}(f)+S_\mathrm{oms}(f)\Big],\\
N_{\zeta_2}(f)&=4\sin^2(\pi fL)\,N_{\zeta}(f).
\end{align}
Here, $S_\mathrm{oms}(f)$ and $S_\mathrm{acc}(f)$ denote the PSDs of the OMS and acceleration noises, respectively. For LISA, TianQin and Taji, these noise PSDs are given by \cite{Babak:2021mhe}
\begin{equation}
\begin{aligned}
S_\mathrm{oms}(f) &= \left(2\pi f P_\mathrm{oms}/c\right)^2\left[1+\left(\frac{2\times10^{-3}\: \mathrm{Hz}}{f}\right)^4\right] \: \mathrm{Hz}^{-1} \, ,\\
S_\mathrm{acc}(f) &= \left(\frac{P_\mathrm{acc}}{2\pi f c}\right)^2\left[1+\left(\frac{0.4\times10^{-3} \: \mathrm{Hz}}{f}\right)^2\right]\,\left[1+\left(\frac{f}{8\times10^{-3} \: \mathrm{Hz}}\right)^4\right]\: \mathrm{Hz}^{-1} \, ,
\end{aligned} 
\end{equation}
where $P_\mathrm{oms}$ and $P_\mathrm{acc}$ are the noise amplitude parameters. For the BBO, the noises PSDs are given by \cite{Corbin:2005ny}
\begin{equation}
\begin{aligned}
    S_\mathrm{oms}^\mathrm{BBO}(f)&=\frac{2.0\times10^{-34}}{3L^2}\:\mathrm{m^2\cdot Hz^{-1}},\\
    S_\mathrm{acc}^\mathrm{BBO}(f)&=\frac{4.5\times10^{-34}}{(2\pi f)^4(3L)^2}\:\mathrm{m^2\cdot Hz^4\cdot Hz^{-1}}.
\end{aligned}
\end{equation}

\begin{table}[h]
    \centering
    \renewcommand{\arraystretch}{1.2}
    \begin{tabular}{lcccc}
        \hline\hline
        \textbf{Parameter} & \textbf{LISA} & \textbf{Taiji} & \textbf{TianQin} & \textbf{BBO} \\
        \hline
        $\text{Arm length}~L~(10^9\mathrm{m})$ & 2.5 & 3 & 0.17 & 0.05 \\
        $P_\mathrm{oms}~(10^{-12}\mathrm{m})$ & 15 & 8 & 1 & / \\
        $P_\mathrm{acc}~(10^{-15}\mathrm{m}\cdot\mathrm{s}^{-2})$ & 3 & 3 & 1 & / \\
        Laser frequency $\nu_0~(\mathrm{Hz})$ & $2.82 \times 10^{14}$ & $2.82 \times 10^{14}$ & $2.82 \times 10^{14}$ & $6 \times 10^{14}$ \\
        Integration time $T_{\mathrm{obs}}~(\mathrm{year})$ & $4.5$ & $5$ & $5/2$ & $4$ \\
        Frequency band $(\mathrm{Hz})$ & $[10^{-4}, 1]$ & $[10^{-4}, 1]$ & $[10^{-4}, 1]$ & $[10^{-1}, 10]$ \\
        \hline\hline
    \end{tabular}
    \caption{Parameters of of several planned SGWDs \cite{LISA:2024hlh,Luo:2019zal,TianQin:2015yph,Crowder:2005nr}.}
    \label{tab:parameters}
\end{table}

\begin{figure}[htbp]
    \centering
    \begin{subfigure}[b]{0.48\textwidth}
        \includegraphics[width=\textwidth]{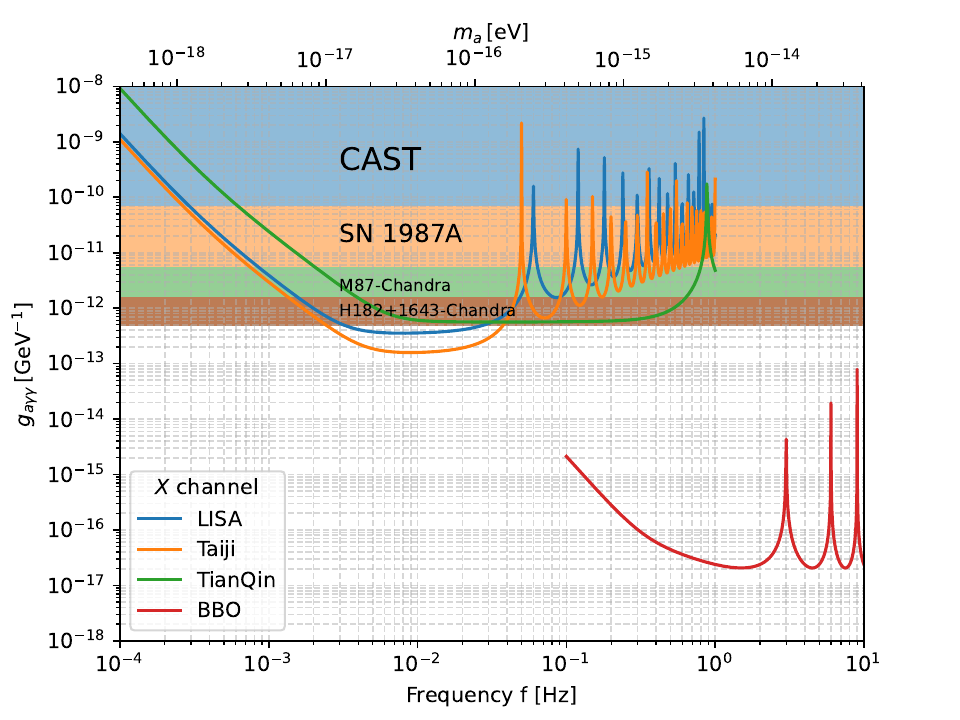}
        \caption{X channel}
        \label{fig:x_channel}
    \end{subfigure}
    \hfill
    \begin{subfigure}[b]{0.48\textwidth}
        \includegraphics[width=\textwidth]{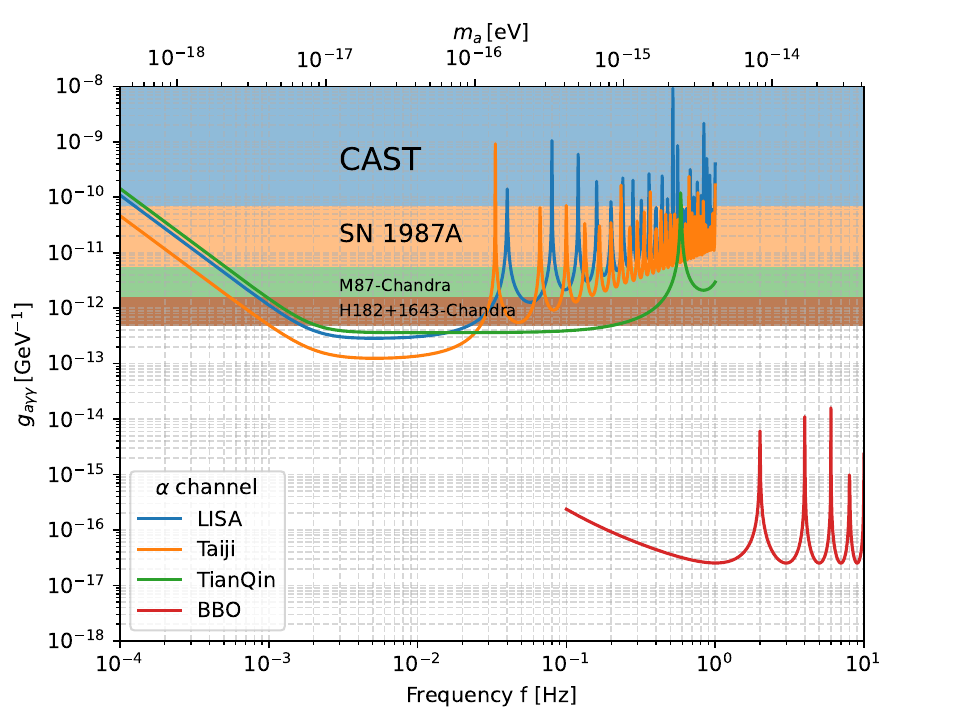}
        \caption{$\alpha$ channel}
        \label{fig:alpha_channel}
    \end{subfigure}\\
    \begin{subfigure}[b]{0.48\textwidth}
        \includegraphics[width=\textwidth]{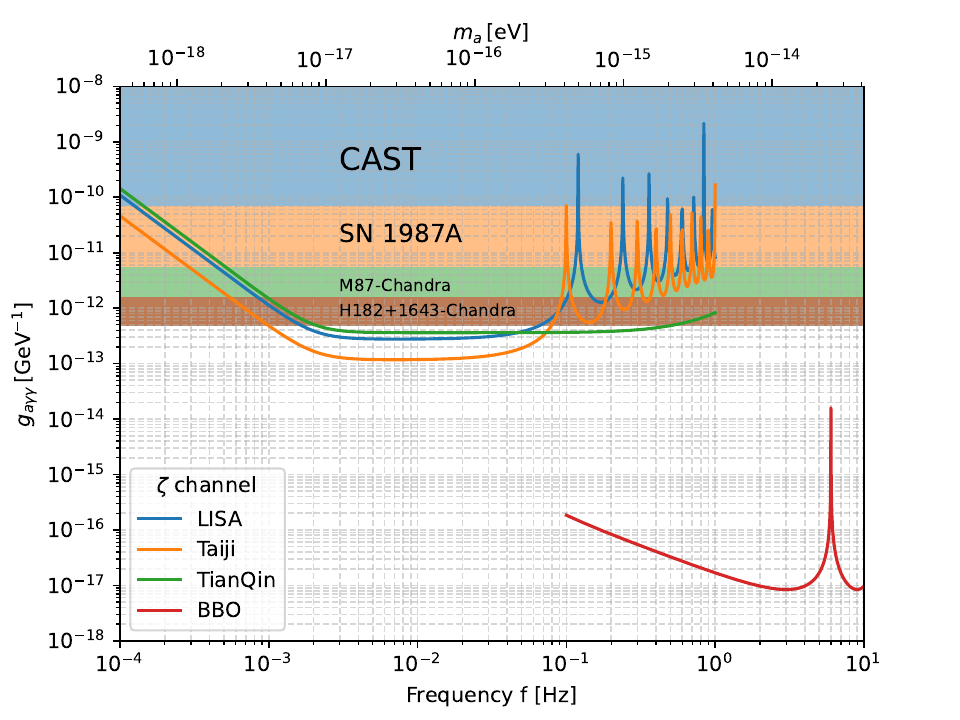}
        \caption{$\zeta$ channel}
        \label{fig:zeta_channel}
    \end{subfigure}
    \caption{Sensitivity estimates of various SGWDs to the ALDM-photon coupling $\gag$ for the first-generation TDI combinations. The constraints from CAST \cite{CAST:2017uph}, SN 1987A \cite{Payez:2014xsa}, the M87 galaxy \cite{Marsh:2017yvc} and the quasar H1821+643 \cite{Reynes:2021bpe} are also shown for comparison.}
    \label{fig:sensitivity}
\end{figure}

Assuming that all optical benches operate with the same laser frequency $\omega_r=2\pi\nu_0$, the sensitivity to the ALDM coupling can be estimated from the signal PSD as
\begin{equation}
    \gag^i(f) = \sqrt{\frac{S_i(f)\, \pi^2 \nu_0^2}{\rho_\mathrm{DM}\, C_i^2(f)\, T_{\mathrm{obs}}}}
\end{equation}
The signal-to-noise ratio (SNR) for each channel is defined by
\begin{equation}
\mathrm{SNR}=\frac{S_i}{N_i}.
\end{equation}
We assume an SNR of 1 to provide a conservative estimate of the interferometer's performance. It is noteworthy that the ratio of noise PSD between the first- and second-generation TDI combinations is the same as the ratio of the squares of their corresponding factors $C_i(f)$. Therefore, the second-generation TDI combinations ($X_2,\,\alpha_2,\zeta_2$) yield sensitivity estimates that are identical to those of their first-generation counterparts. In Fig. \ref{fig:sensitivity}, we illustrate the sensitivity estimates of LISA, Taiji, TianQin, and BBO to the ALDM-photon coupling $\gag$ for the first-generation TDI combinations. These estimates are compared with existing constraints derived from CAST \cite{CAST:2017uph}, SN 1987A \cite{Payez:2014xsa}, the M87 galaxy \cite{Marsh:2017yvc} and the quasar H1821+643 \cite{Reynes:2021bpe}. The parameters of interest for the sensitivity curves of various SGWDs are summarized in Table. \ref{tab:parameters}.

The comparisons to existing astrophysical and laboratory constraints reveal that the SGWD-based approach holds potential to improve sensitivities to the ALDM-photon coupling strength, particularly in the mass range of $10^{-19}\,\mathrm{eV}$ to $10^{-14}\,\mathrm{eV}$. Notably, the SGWD sensitivity curves dip below the astrophysical constraints in certain mass windows, reaching $\gag\sim10^{-13}\:\mathrm{GeV^{-1}}$, signifying their competitive reach in the search for ALDM. 

The sensitivity curves for various experimental configurations and TDI channels demonstrate that these detectors can probe a wide range of ALDM masses. The inclusion of different SGWDs shows their complementary coverage across the ALDM parameter space. The BBO extends sensitivity toward higher-frequency domains, covering a distinct region of parameter space. The sensitivity curves also indicate that certain TDI combinations offer superior detection potential for specific mass ranges. In particular, the $\zeta$ channel exhibits better performance across a wide range of ALDM masses.

Future work could refine these projections by incorporating potential noise sources and optimizing interferometric configurations to further enhance detection capabilities.

\section{\label{sec:discussion}Discussion}

The proposed OB modification offers a promising route to probe the ALDM-photon coupling without altering the polarization state of inter-spacecraft links. However, the practical implementation of this modification entails a series of intricate engineering challenges that demand meticulous consideration. For instance, the stabilization of additional optical elements is critical to avoid introducing extra phase noise. Specifically, the 50/50 beam splitter ratio for the AUXI must be carefully maintained to prevent noise arising from power fluctuations, while additional baffling for the auxiliary optical path is required to mitigate stray light effects.

The projected sensitivity curves, presented in the preceding section, provide a preliminary assessment of the instrument's performance, demonstrating the potential of SGWDs to probe ALDM. Despite these promising sensitivity projections, the impact of the modification on the overall OB design cannot be overlooked. The integration of an auxiliary interferometer into the OB introduces challenges not only in maintaining mechanical and thermal stability but also in managing the total light power budget. Systematic errors arising from the auxiliary path, such as those induced by imperfect beam splitting or misalignment, must be carefully quantified and minimized. The p-polarized light induced by the ALDM may be mitigated by the existing optics of the OB like transmitting beam clipping, requiring dedicated design for the actual light path. Future research should focus on detailed simulations and laboratory demonstrations to validate the preliminary sensitivity estimates and to develop practical methods for mitigating the additional noise contributions introduced by the modification.

In summary, while the analytical projections and sensitivity curves are encouraging, the operational realization of the proposed OB modification in SGWDs necessitates the resolution of significant engineering hurdles. The feasibility of this approach hinges on the successful integration of the auxiliary interferometer without compromising the performance of the primary gravitational wave detection channels. If these challenges can be effectively addressed, the modified OB design could emerge as a powerful new tool in the search for ALDM, thereby expanding the scientific capabilities of SGWDs.

\section{\label{sec:conclusion}Conclusion}

In this work, we have proposed a modification to the OB in SGWDs to detect ALDM. We presented an alternative strategy by introducing an auxiliary interferometer that converts ALDM-induced polarization rotations into measurable phase shifts, without changing the polarization of the inter-spacecraft laser light. The projected sensitivity to the ALDM-photon coupling covers a broad range of ALDM masses, from $10^{-19}\,\mathrm{eV}$ to $10^{-14}\,\mathrm{eV}$. In general, SGWDs such as LISA, Taiji, and TianQin could achieve sensitivities of $\gag\sim10^{-13}\:\mathrm{GeV^{-1}}$ for ALDM with masses around $10^{-17}\:\mathrm{eV}$. A combined analysis of all TDI channels could further improve the reach. This approach preserves the design integrity of current SGWD optical benches while significantly expanding their scientific capability to explore new physics.

\begin{acknowledgments} 
The authors would like to thank the anonymous reviewers for their insightful comments and constructive suggestions, which helped to improve the quality of this paper.
QGH is supported by the National Key Research and Development Program of China Grant No.2020YFC2201502, grants from NSFC (grant No. 12475065, 12250010).
RMY is supported by the National Natural Science Foundation of China under Grant No. 12447131.
BXJ is supported by the National Natural Science Foundation of China under Grant No. 12447105.
\end{acknowledgments}


\bibliographystyle{JHEP}
\bibliography{ALDM}

\end{document}